\documentclass[prd,aps, tightenlines, preprintnumbers, showpacs, nofootinbib,superscriptaddress,notitlepage,11pt]{revtex4-1}

\usepackage{graphicx}
\usepackage{amsmath,amsthm,amssymb}
\usepackage{mathtools}
\usepackage{physics}
\usepackage{siunitx}

\newcommand\dA{\ensuremath{\mathrm{d}A}}
\newcommand\pA{\ensuremath{\mathrm{p}A}}
\newcommand\pPb{\ensuremath{\mathrm{pPb}}}
\newcommand\dAu{\ensuremath{\mathrm{dAu}}}
\newcommand\pp{\ensuremath{\mathrm{pp}}}
\newcommand\hardscale{\ensuremath{Q}}
\newcommand\satscale{\ensuremath{Q_s}}
\newcommand\xprojectile{\ensuremath{x_p}}
\newcommand\xtarget{\ensuremath{x_g}}
\newcommand\zhadron{\ensuremath{z}}
\newcommand\massnumber{\ensuremath{A}}
\newcommand\lperp{\ensuremath{l_\perp}}
\newcommand\lpperp{\ensuremath{l_\perp'}}
\newcommand\lppperp{\ensuremath{l_\perp''}}
\newcommand\kperp{\ensuremath{k_\perp}}
\newcommand\pperp{\ensuremath{p_\perp}}
\newcommand\qperp{\ensuremath{q_\perp}}
\newcommand\uperp{\ensuremath{u_\perp}}
\newcommand\upperp{\ensuremath{u_\perp'}}
\newcommand\umin{\ensuremath{u_<}}
\newcommand\bperp{\ensuremath{b_\perp}}
\newcommand\xperp{\ensuremath{x_\perp}}
\newcommand\yperp{\ensuremath{y_\perp}}

\newcommand\rperp{\ensuremath{r_\perp}}
\newcommand\Rperp{\ensuremath{R_\perp}}
\newcommand\sqsnn{\ensuremath{\sqrt{s_{NN}}}}
\newcommand\pseudorapidity{\ensuremath{\eta}}
\newcommand\rapidity{\ensuremath{y}}
\newcommand\alphas{\ensuremath{\alpha_s}}
\newcommand\Nc{\ensuremath{N_c}}
\newcommand\centrality{\ensuremath{c}}
\newcommand\targetarea{\ensuremath{S_\perp}}
\newcommand\RpA{\ensuremath{R_{\pA}}}
\newcommand\Ncoll{\ensuremath{N_{\text{coll}}}}
\newcommand\besselJzero{\ensuremath{J_0}}
\newcommand\besselIzero{\ensuremath{I_0}}
\newcommand\defn{\equiv}
\newcommand\etal{\textit{et al}}

\begin{document}
\title{Implementing the exact kinematical constraint in the saturation formalism}

\author{Kazuhiro Watanabe}
\affiliation{Key Laboratory of Quark and Lepton Physics (MOE) and Institute
of Particle Physics, Central China Normal University, Wuhan 430079, China}

\author{Bo-Wen Xiao}
\affiliation{Key Laboratory of Quark and Lepton Physics (MOE) and Institute
of Particle Physics, Central China Normal University, Wuhan 430079, China}

\author{Feng Yuan}
\affiliation{Nuclear Science Division, Lawrence Berkeley National
Laboratory, Berkeley, CA 94720, USA}

\author{David Zaslavsky}
\affiliation{Key Laboratory of Quark and Lepton Physics (MOE) and Institute
of Particle Physics, Central China Normal University, Wuhan 430079, China}

\begin{abstract}
We revisit the issue of the large negative next-to-leading order (NLO) cross 
section for single inclusive hadron production in $\pA$ collisions in the 
saturation formalism. By implementing the exact kinematical constraint 
in the modified dipole splitting functions, two additional positive NLO 
correction terms are obtained. In the asymptotic large $\kperp$ limit, we 
analytically show that these two terms become as large 
as the negative NLO contributions found in our previous calculation. 
Furthermore, the numerical results demonstrate that the applicable regime 
of the saturation formalism can be extended to a larger $\kperp$ window, 
where the exact matching between the saturation formalism (in the asymptotic $\kperp$ 
regime) and the collinear factorization calculations will have to be performed separately. 
In addition, after significantly improving the numerical accuracy of the NLO correction, we obtain 
excellent agreement with the LHC and RHIC data for forward hadron productions. 
\end{abstract}
\maketitle

\section{Introduction}
A focal point of the frontier of high energy nuclear physics at RHIC and the LHC is the study of saturation in hadron collisions. Saturation is an effect that emerges due to bremsstrahlung gluon radiation in the hadronic wavefunction. It was prompted by the theoretical prediction that, at high energy, the gluon density rises rapidly as $x$, the longitudinal momentum fraction of the gluons with respect to their parent hadron, decreases. This rise is governed by the famous Balitskii, Fadin, Kuraev, and Lipatov (BFKL) evolution equation~\cite{BFKL}, which emerges from resummation of terms proportional to $\alphas \ln\frac{1}{x}$. However, when the gluon density becomes high, it is expected that gluons start to recombine and QCD dynamics becomes nonlinear. This eventually leads to the onset of gluon saturation~\cite{Gribov:1984tu, Mueller:1985wy, McLerran:1993ni}, as a result of the nonlinear QCD dynamics.
%Saturation physics provides us detailed description of dense parton densities in high energy limit.
To quantify the recombination effect, a nonlinear term in the gluon evolution equation was proposed in Ref.~\cite{Gribov:1984tu, Mueller:1985wy}. This nonlinear extension of the BFKL evolution equation was later independently derived by Balitsky and Kovchegov; accordingly, the equation is referred to as the BK evolution equation~\cite{Balitsky:1995ub, Kovchegov:1999yj}. Theoretically, it seems that gluon saturation is bound to occur as a result of high energy evolution.

The task remains to find a ``smoking gun'' signature of gluon saturation in \emph{experimental} data at e.g. RHIC or the LHC. A wealth of results in $\pA$ collisions, ideal for observing saturation~\cite{Salgado:2011pf}, are becoming available. But it is critical to have precise, quantitative phenomenological calculations in the saturation formalism to compare to these experimental results.

Single inclusive hadron production in $\pA$ collisions at high energy reveals the interesting physics of gluon saturation particularly well, compared to $\pp$ collisions. The effect of dense gluons in the target nucleus can be characterised by the introduction of a semi-hard momentum scale, which is known as the saturation scale $\satscale$, a function of the momentum fraction $\xtarget = \frac{k_{\text{gluon}}^+}{k_{\text{nucleon}}^+}$, and the nuclear mass number $A$. Roughly speaking, the saturation scale can be used to separate the saturated (dense) regime, in which the nonlinear energy evolution applies, from the (dilute) regime in which the evoution is linear. When the typical hard scale of the scattering, $\hardscale$, is less than $\satscale$, one expects that the target partons involved in the interaction are saturated. On the other hand, when $\hardscale \gg \satscale$,  the saturation effect is no longer important, and standard perturbative QCD should be sufficient to describe the data. 

It is generally believed that the transverse momentum of typical gluons inside nuclear targets is roughly $\satscale(\xtarget, \massnumber)$. In high energy $\pA$ collisions, before partons from the proton projectile fragment into hadrons with transverse momentum $\pperp$, they undergo multiple interactions with the dense gluonic fields in the highly boosted target nucleus, picking up a transverse momentum of roughly $\satscale$ in the process. The squared saturation momentum $\satscale^2$ in nuclear targets is $\massnumber^{1/3}$ times of that in protons, due to random multiple scatterings. Therefore, the transverse momentum ($\pperp$) spectrum of the produced hadrons exhibits different behaviour in $\pA$ collisions, especially in the relatively low $\pperp$ regime, than in $\pp$ collisions.

Measurements of single inclusive hadron production in $\pA$ collisions at RHIC~\cite{Adler:2003kg, Arsene:2004ux, Adams:2003qm, Adams:2006uz, Braidot:2010ig, Adare:2011sc} 
and the LHC~\cite{ALICE:2012xs, ALICE:2012mj, Abelev:2014dsa, Hadjidakis:2011zz, LHCb, CMS:2013cka, Atlas, AlexanderMilovonbehalfoftheATLAS:2014rta} have provided plenty of data.
There have been many theoretical and phenomenological efforts~\cite{Dumitru:2002qt,Albacete:2013ei, Dumitru:2005kb, Albacete:2010bs, Levin:2010dw, Altinoluk:2011qy, Fujii:2011fh, Tribedy:2011aa,  Albacete:2012xq, Xu:2012au, Kang:2012kc, Basso:2012nb, Rezaeian:2012ye, Lappi:2013zma,  Deng:2014vda, Albacete:2014fwa, Albacete:2013tpa,  Kutak:2014wga} on this subject, which imply that gluon saturation (or shadowing) plays an important role in the production of forward rapidity hadrons.

The first complete next-to-leading order (NLO) calculation for inclusive hadron productions~\cite{Chirilli:2011km, Chirilli:2012jd} in $\pA$ collisions was achieved a few years ago, using the Mueller's dipole formalism (see Ref~\cite{Mueller:2012bn} for discussion at NLO). This computation is the first complete NLO calculation which uses dimensional regularization and with the $\overline{\textrm{MS}}$ regularization scheme, which is necessary to correctly incorporate the available NLO collinear parton distributions (PDFs) and fragmentation functions (FFs) without introducing additional scheme dependence.

Schematically, the full NLO cross section for hadron production at forward rapidity $\rapidity$, with the hadron having transverse momentum $\pperp = \zhadron\kperp$, can be written as follows:
\begin{equation}
\frac{\dd[3]\sigma}{\dd \rapidity \dd[2]\pperp} = \int \xprojectile f_i(\xprojectile)\otimes D_{h/i}(\zhadron) \otimes \mathcal{F}^{\xtarget}_i(\kperp) \otimes \mathcal{H}^{(0)} 
+\frac{\alphas}{2\pi} \int \xprojectile f_i(\xprojectile)\otimes D_{h/j}(\zhadron) \otimes \mathcal{F}_{(N)ij}^{\xtarget}\otimes \mathcal{H}_{ij}^{(1)},
\end{equation}
All the hard factors $\mathcal{H}$ are given in Refs.~\cite{Chirilli:2011km, Chirilli:2012jd}. $\mathcal{F}^{\xtarget}_i(\kperp)$ and $\mathcal{F}_{(N)ij}^{\xtarget}$, which are functions of the transverse momentum $\kperp$ of the produced parton, represent the Fourier transforms of dipole scattering amplitudes. $\xprojectile f_i(\xprojectile)$ and $D_{h/j}(\zhadron)$ are the parton distribution and fragmentation function, respectively. 

The first \emph{numerical} analysis of forward hadron production in $\pA$ and $\dA$ collisions in the small-$x$ saturation formalism, based on the NLO results in Refs.~\cite{Chirilli:2011km, Chirilli:2012jd}, was presented in Refs.~\cite{Stasto:2013cha, Zaslavsky:2014asa}. It was found that the theoretical uncertainty is significantly reduced compared to leading order (LO) results, and the calculated NLO cross section agrees well with forward-rapidity RHIC data for $\pperp \lesssim \satscale$. Recall that $\satscale$ is the characteristic scale for the gluon density in a heavy nucleus. In general, the $\pperp$ region in which the calculation and results agree increases with the center-of-mass scattering energy $\sqsnn$, since the typical gluon density probed is larger at high energy. However, the numerical results of the NLO cross section abruptly drop to negative values above some cutoff momentum which is generally slightly greater than $\satscale$.\footnote{The relationship between the saturation scale $\satscale$, the cutoff momentum at which the results become negative, and the boundary of the region in which the calculation accurately describes the data appears to be some sort of rapidity-dependent proportionality, but the details are not clear. Ref.~\cite{Zaslavsky:2014asa} includes some discussion of the relationship among these momenta.}

Strictly speaking, the saturation formalism always takes the high energy limit 
$s\to \infty$, which yields large saturation momentum $\satscale$. In this limit, 
the NLO results in Refs.~\cite{Chirilli:2011km, Chirilli:2012jd} are obtained after 
the subtraction of the rapidity divergence (associated with the small-$x$ evolution 
in the $s\to \infty$ limit) as well as the collinear divergences (associated with the 
DGLAP evolution of PDFs and FFs). However, in phenomenological studies of 
saturation physics, the center-of-mass energy of scatterings is finite and the 
saturation momentum is not very large. It is natural to expect that the saturation 
formalism works when $\kperp \leq \satscale$. On the other hand, when 
$\kperp \gg \satscale$, the saturation formalism is believed to be no longer 
applicable since $\xtarget$ is no longer small, and the collinear factorization 
approach should be the relevant formalism to describe the large $\kperp$ part 
of the cross section. Studies~\cite{Dominguez:2010xd} have shown that the 
transverse momentum dependent (TMD) factorization is closely related or 
equivalent to the small-$x$ factorization in terms of gauge links~\cite{Boer:2003cm, Bomhof:2006dp, Xiao:2010sp}, 
if one puts them in the same kinematical region. 
A similar kinematic restriction is also required for the TMD factorization in the
hard scattering processes, where the transverse momentum $k_\perp$ is much smaller than
the hard momentum scale $Q$ (like the invariant mass of lepton pair
in the Drell-Yan process)~\cite{Collins:2011zzd}. In other words, the TMD
factorization would be invalid in the large $k_\perp\sim Q$ region. In this region,
a matching to the collinear factorization calculation is usually performed~\cite{Collins:2011zzd}.
This argument applies to the case studied in this paper as well.

In Ref.~\cite{Stasto:2014sea}, the matching between the perturbative results is reached by taking the large $\kperp$ limit of the NLO cross section in the small-$x$ formalism and enforcing the exact kinematic constraint. One might wonder if there is a way to naturally implement the kinematic constraint in the small-$x$ formalism. Doing so could help extend the applicability of the formalism to the large $\kperp$ regime. Recently, the authors of Refs.~\cite{Beuf:2014uia, Altinoluk:2014eka} performed another independent NLO calculation of single inclusive hadron production in $\pA$ collisions. Their discussion of the Ioffe time dependence, which is equivalent to the kinematical constraint, motivated us to investigate the details of the dipole formalism application in this process
and compute the effect of the kinematical constraints. We find that we obtain two additional NLO corrections from incorporating these constraints. These two terms were conjectured to be small in the $s\to \infty$ high energy limit, and therefore implicitly neglected in the original derivation of the NLO corrections~\cite{Chirilli:2011km, Chirilli:2012jd}.
It is interesting to note that, in Ref.~\cite{Mueller:2013wwa}, a similar logarithmic term played an important role in deriving the Sudakov factor in other hard processes. We need to emphasize that there is no Sudakov factor in the process of single hadron productions. As we will show later, these additional NLO corrections are indeed small as long as $\kperp < \satscale$. However, they become large when $\kperp > \satscale$. By including these two terms, we can offset the negativity of the NLO terms at larger $\pperp$ and extend the applicability window of the saturation formalism towards larger $\pperp$ for single hadron productions. 

The goal of this paper is to revisit the issue of the negative NLO cross section found in the large $\kperp$ regime of forward-rapidity single inclusive hadron production in $\pA$ collisions, and the numerical implementation of the exact kinematics in the small-$x$ formalism at one-loop order. We find that, with the exact kinematical constraint imposed, we can obtain two additional NLO hard factors, with one from the $q\to q$ channel and the other from the $g\to g$ channel. We first analytically show that these two terms are large enough to partially overcome the negative NLO terms found earlier in the simple Golec-Biernat and Wusthoff (GBW) model~\cite{GolecBiernat:1998js}. Numerically, these two terms are found to be negligible when $\kperp < \satscale$, but they become important when $\kperp$ rises to $\approx 2 \satscale$ and higher. 

More importantly, we have significantly improved our numerical implementation of all NLO corrections, which allows us to do phenomenological studies at the LHC energy. We find excellent agreement between the full NLO cross section and the forward hadron production data at the LHC. This paves the road for a quantitative and precise phenomenological test of saturation physics at the LHC. 

The rest of this paper is organized as follows. In Sec. II, we present a detailed derivation of the implementation of the kinematical constraint into the dipole model, and obtain two additional NLO corrections (one for the quark channel and one for the gluon channel) after subtracting the corresponding small-$x$ large logarithms. 
We further evaluate these two terms by Fourier transform, and demonstrate that they have the same $\frac{1}{\kperp^4}$ behavior exhibited by perturbative QCD in the high $\kperp$ limit.  In particular, we analytically show that the additional terms are large compared to the negative cross section at NLO. In Sec. III, we present the numerical results and compare them to RHIC and LHC results. 
We summarize our paper in Sec. IV.

\section{Dipole model and Kinematical Constraints at NLO}
 
\begin{figure}[tbp]
 \centering
 \includegraphics[height=4cm]{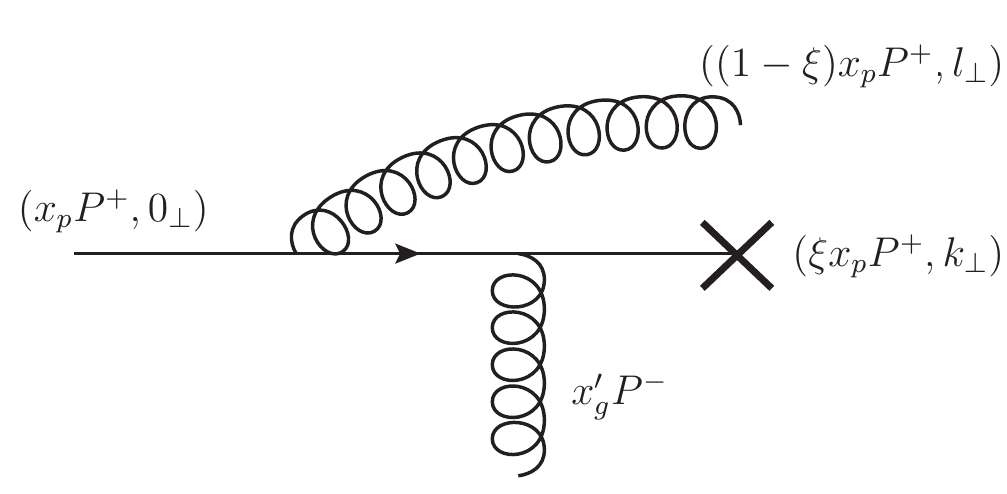} 
 \caption[*]{A typical real diagram at NLO}
 \label{ki}
\end{figure}

Let us first review the exact kinematical constraint discussed in previous publications~\cite{Mueller:2013wwa, Stasto:2014sea, Beuf:2014uia,  Altinoluk:2014eka}, since it plays an important role in this paper. We will continue to use light-cone perturbation theory as in Refs.~\cite{Chirilli:2011km, Chirilli:2012jd}, and define $p^+=\frac{p^0+p^3}{\sqrt{2}}$ and $p^-=\frac{p^0-p^3}{\sqrt{2}}$.

The kinematical constraint is derived from the conservation of the $-$ component of the four momentum before and after interactions for $2\to2$ processes. For quark production with transverse momentum $\kperp$, as illustrated in Fig.~\ref{ki}, we obtain
\begin{equation}\label{eq:kinematicalconstraint}
\xtarget^\prime P^- =\frac{\lperp^2}{2(1-\xi)\xprojectile P^+}+\frac{\kperp^2}{2\xi \xprojectile P^+} \leq P^- .
\end{equation}
As $\xi$ approaches $1$, the above kinematical constraint indicates
\begin{equation}
\lperp^2\leq (1-\xi)\xprojectile s \, ,\quad \textrm{or}\quad \xi \leq 1-\frac{\lperp^2}{\xprojectile s}. \label{con}
\end{equation}                                                                         
For a rapidity divergent term, we find that the above constraint modifies the upper limit of the divergent integral as follows
\begin{equation}
\int_0^{1-\frac{\lperp^2}{\xprojectile s}} \frac{\dd\xi}{1-\xi} =\ln\frac{\xprojectile s}{\lperp^2} =\ln\frac{1}{x_g} +\ln\frac{\kperp^2}{\lperp^2}, \label{mki}
\end{equation}
where $x_g \equiv \frac{\kperp^2}{\xprojectile s}$ according to the leading order kinematics. In the high energy limit, we assume $\kperp \sim \lperp$ which makes the second term very small. However, as we shall show in the following discussion, the $\ln\frac{\kperp^2}{\lperp^2}$ term becomes important when $\kperp$ gets larger than the typical saturation momentum. The direct evaluation of this $\ln\frac{\kperp^2}{\lperp^2}$ term is not easy in the momentum space, but indirectly evaluating it in coordinate space is quite straightforward, as we shall demonstrate in the following discussion, after encoding it into the modified dipole splitting functions.%
\footnote{A similar term, $\ln \frac{M^2}{\lperp^2}$, where $M$ is another hard scale (e.g. the Higgs mass or dijet invariant mass), is used to derive the Sudakov factor when $M^2\gg \kperp^2$~\cite{Mueller:2013wwa}. The Sudakov physics is different from what we are discussing here, since basically the single hadron production process is a single scale problem. 
First of all, the new terms that we obtain can be never interpreted as a Sudakov factor, instead it should be viewed as part of the NLO power correction. In addition, the color flow is completely different. Note that the color factor for the quark Sudakov factor is $C_F$ while the one for BFKL physics is always $N_c/2$. }

Inspired by Refs.~\cite{Beuf:2014uia, Altinoluk:2014eka}, we find that it is convenient to encode the kinematical constraint by modifying the dipole splitting function for the $q\to qg$ splitting, shown in Fig.~\ref{ki}, as follows:
\begin{equation}
\psi^\lambda_{\alpha\beta}(p^+,k^+,u_{\perp})=2\pi i\sqrt{\frac{2}{(1-\xi)p^+}} \left[1-\besselJzero\left(\uperp \Delta \right)\right]
\begin{cases}
\frac{u_{\perp}\cdot\epsilon^{(1)}_\perp}{u_{\perp}^2}(\delta_{\alpha-}%
\delta_{\beta-}+\xi\delta_{\alpha+}\delta_{\beta+}), & \lambda=1, \\
\frac{u_{\perp}\cdot\epsilon^{(2)}_\perp}{u_{\perp}^2}(\delta_{\alpha+}%
\delta_{\beta+}+\xi\delta_{\alpha-}\delta_{\beta-}), & \lambda=2.%
\end{cases}
\ ,  \label{wfun}
\end{equation}
with $\Delta^2=\xi(1-\xi)\xprojectile s$. The original dipole splitting function, which is proportional to $\frac{u_{\perp}\cdot\epsilon^{(1,2)}_\perp}{u_{\perp}^2}$, arises from the Fourier transform of $\frac{\lperp\cdot\epsilon^{(1,2)}_\perp}{\lperp^2}$. The additional term of $-\besselJzero(\uperp\Delta)$ arises from the kinematical constraint, Eq.~\eqref{con}, imposed during the Fourier transform.

In general, the correction to the splitting function $\besselJzero\left(\uperp \Delta \right)$ does not play any important role since it vanishes when we take the high energy limit $s\to \infty$. It only becomes important when the gluon longitudinal momentum fraction $1-\xi$ approaches zero. Specifically, for quark production at one-loop order, we always get the following DGLAP-type contribution from real diagrams:
\begin{equation}
\int_\tau^1 \dd\xi \frac{1+\xi^2}{1-\xi} =\int_\tau^1 \dd\xi \frac{1+\xi^2}{(1-\xi)_+} +\int_0^1 \dd\xi \frac{2}{1-\xi}.
\end{equation}
For the first term on the right hand side of the above equation, we can safely take the $s\to \infty$ limit, since this term is regular when $\xi \to 1$. However, one can not neglect the correction $\besselJzero\left(\uperp \Delta \right)$ for the second term, since it is singular when $\xi\to 1$. Clearly, in this NLO calculation for single hadron production in $\pA$ collisions, the kinematical constraint only affects the rapidity subtraction term.

The relevant contribution to the cross section, with the modified splitting function, can be written as 
\begin{multline}\label{eq:contribution}
\frac{\alphas \Nc}{2\pi ^{2}} \int_{0}^{1}\frac{\dd\xi }{1-\xi }
\int \frac{\dd[2]\xperp \dd[2]\yperp \dd[2]\bperp}{(2\pi )^{2}} e^{-i\kperp\cdot (\xperp-\yperp)}
\Bigl[-S(\xperp,\yperp) + S(\xperp,\bperp)S(\bperp,\yperp)\Bigr] \\
\times\biggl\{\frac{\bigl[1-\besselJzero(\uperp\Delta)\bigr]^2}{\uperp^2}
+ \frac{\bigl[1-\besselJzero(\upperp\Delta)\bigr]^2}{\upperp^2}-\frac{2\uperp\cdot\upperp}{\uperp^2 \upperp^2}\bigl[1-\besselJzero(\uperp\Delta)\bigr]
\bigl[1-\besselJzero(\upperp\Delta)\bigr]\biggr\}, 
\end{multline}
where $\uperp \defn \xperp - \bperp$ and $\upperp \defn \yperp - \bperp$.%
\footnote{This term looks similar to the last term of Eq.~(4.21) in Ref.~\cite{Altinoluk:2014eka}. Here we are implementing the kinematical constraint, while they are discussing the Ioffe time cutoff. These two become equivalent if one identifies their $\frac{2P^+}{\tau}$ as the center-of-mass energy $s$ in our paper.}
One can actually approximately integrate over $\xi$ and find that%
%\footnote{}
\begin{gather}
\int_0^{1}\frac{\dd \xi}{1-\xi} \biggl[1-\besselJzero\Bigl(\uperp \sqrt{\xprojectile s  (1-\xi})\Bigr)\biggr]^2
\simeq \ln\frac{\xprojectile s \uperp^2}{c_0^2} =\ln \frac{1}{\xtarget}+\ln\frac{\kperp^2 \uperp^2}{c_0^2} \\
\begin{multlined}[\textwidth-4em]
\int_0^{1}\frac{\dd \xi}{1-\xi} \biggl[1-\besselJzero\Bigl(\uperp \sqrt{\xprojectile s (1-\xi})\Bigr)\biggr] \biggl[1-\besselJzero\Bigl(\upperp \sqrt{\xprojectile s (1-\xi})\Bigr)\biggr]\\ 
\simeq  \ln\frac{\xprojectile s \uperp\upperp}{c_0^2}=\ln \frac{1}{\xtarget}+\ln\frac{\kperp^2 \uperp \upperp}{c_0^2}\, ,
\end{multlined}
\end{gather}
with $c_0=2e^{-\gamma_E}$. Here we have used $\xprojectile \xtarget s = \kperp^2$. It is then clear that the first term $\ln\frac{1}{\xtarget}$ can be subtracted from the NLO cross section and 
interpreted as the BK evolution of the dipole amplitude $S$ up to rapidity $Y_g=\ln\frac{1}{\xtarget}$. The second term in the above equations, which is conjugate to the term $\ln\frac{\kperp^2}{\lperp^2}$ as in Eq.~\eqref{mki} (see also Eq.~(3.12) of Ref.~\cite{Altinoluk:2014eka}) with $\lperp$ being the gluon transverse momentum, arises due to the exact kinematical constraint. More precisely, the second integral should give $\ln \frac{1}{\xtarget}+\ln\frac{\kperp^2 \umin^2}{c_0^2}$ instead of $\ln \frac{1}{x_g}+\ln\frac{\kperp^2 \uperp \upperp}{c_0^2}$ with $\umin \defn \operatorname{min}\{\uperp, \upperp\}$, which makes the calculation for $L_q(\kperp)$ non-analytical and the precise numerical evaluation more challenging. Fortunately, one can numerically check that the resulting $L_q(\kperp)$ has similar large-$\kperp$ behaviour, and it gives the same high $\kperp$ tail, since $\uperp\simeq \upperp$ when $\kperp \to \infty$. Besides, as we will show later, in the low-$\kperp$ region, $L_q(\kperp)$ is negligible in the total cross section. Our goal here is to extract the correct large $\kperp$ tail of the additional hard factor, which eventually helps to extend the applicability of the small-$x$ calculation. In this sense, we can approximate $\ln\frac{\kperp^2 \umin^2}{c_0^2}$ as $\ln\frac{\kperp^2 \uperp\upperp}{c_0^2}$. Also, because the rest of the expression is symmetric under the exchange $\uperp \leftrightarrow \upperp$, it is useful to note that this is equivalent to using $\ln\frac{\kperp^2 \uperp^2 }{c_0^2}$ or $\ln\frac{\kperp^2 \upperp^2}{c_0^2}$.

The leftover terms can be cast into an additional hard factor which reads
\begin{multline}
L_q (\kperp) = \frac{\alphas\Nc}{2\pi^2} \int \frac{\dd[2]\xperp \dd[2]\yperp \dd[2] \bperp}{(2\pi)^2}
e^{-i\kperp \cdot (\xperp - \yperp)} \bigl[S(\xperp - \bperp) S(\yperp - \bperp) -S(\xperp - \yperp)\bigr] \\
\times  \biggl[\frac{1}{\uperp^2} \ln\frac{\kperp^2 \uperp^2}{c_0^2}+\frac{1}{\upperp^2} \ln\frac{\kperp^2 \upperp^2}{c_0^2} -\frac{2\uperp \cdot \upperp}{\uperp^2 \upperp^2} \ln\frac{\kperp^2 \abs{\uperp}\abs{\upperp}}{c_0^2}\biggr]. \label{lq1}
\end{multline}
The corresponding contribution to the single inclusive cross section in this channel can be written as 
\begin{equation}
\frac{\dd[3] \sigma_{L_q}}{\dd\rapidity\dd[2]\pperp} =\int_\tau^1\frac{\dd z}{z^2} \sum_f \xprojectile q_f(\xprojectile) D_{h/q} (z) L_q(\kperp). \label{sigl}
\end{equation}
It is not hard to show that the above contribution from $L(\kperp)$ is free of both UV and IR divergences. When $\bperp \to \xperp$, the first bracket vanishes. When $\bperp \to \infty$, the second bracket vanishes. Due to these strong cancellations, it was believed that this contribution should be small. In fact, Ewerz \etal~\cite{Ewerz:2009nv} studied the Ioffe time effect of the dipole model in deep inelastic scattering for inclusive total cross sections, and they found that this effect \emph{is} small. For single inclusive hadron production in $\pA$ collisions, as we demonstrate below, the effect is small when $\pperp$ is small, but it becomes as large as other NLO corrections when $\pperp \sim \satscale$. %Thus, this term is neglected in the calculation in Ref.~\cite{Chirilli:2012jd}. 

Note that this term is physically and fundamentally different from the so-called $\Delta H$ correction from Kang \etal~\cite{Kang:2014lha}, which is proportional to the rapidity interval $Y - Y_g=\ln \frac{1}{\xprojectile}+\ln\frac{\kperp^2}{m_p^2}$. As commented in Ref.~\cite{Xiao:2014uba}, the choice of the rapidity interval leads to an unphysical conclusion and violates the small-$x$ factorization. The new additional term $L_q (\kperp)$ does not depend on either the projectile longitudinal momentum fraction $\xprojectile$, or the hadronic mass $m_p$. It is important to notice that QCD factorization does not allow us to have hadronic mass $m_p$ in any hard factors. Otherwise, this implies that we can not separate the non-perturbative physics from the perturbative calculable hard factors. It is also clear from our above derivation that $\xprojectile$ naturally cancels out and thus does not appear in $L_q$. We would like to emphasize that the so-called $\Delta H$ correction discussed in Ref.~\cite{Kang:2014lha} is unjustified and should be absent in view of the small-$x$ factorization.

Let us derive the following simplified expression for $L_q(\kperp)$ which is much easier to evaluate numerically. It is straightforward to use the following Fourier transform identities%
\footnote{Note that this Fourier transform may be problematic for $\uperp=0$, therefore we should exclude the point where $\uperp=0$, in principle. However, since the first bracket in Eq.~\eqref{lq1} vanishes when $\xperp \to \bperp$ (or equivalently $\uperp\to 0$), which suggests that we are justified in ignoring the fact that $L_q(\kperp)$ is undefined at that point. We have also numerically tested that the two expressions of $L(\kperp)$, before and after the Fourier transform, give the same numerical results.}
\begin{align}
\frac{1}{\uperp^2}\ln \frac{\kperp^2 \uperp^2}{c_0^2} &= \frac{1}{8\pi}\int \dd[2]\lperp e^{i\lperp\cdot \uperp} \biggl(\ln\frac{\kperp^2}{\lperp^2}\biggr)^2 \\
\frac{\vec{u}_\perp}{\uperp^2}\ln \frac{\kperp^2 \uperp^2}{c_0^2} &= \frac{1}{2\pi}\int \dd[2]\lperp e^{i\lperp\cdot\uperp} \frac{i\vec{l}_\perp}{\lperp^2}\ln\frac{\kperp^2}{\lperp^2}
\end{align}
to find that
\begin{align}
 L_q (\kperp) &= \frac{\alpha_s N_c}{4\pi^2} S_\perp \left[L_{q1} (\kperp)+L_{q2} (\kperp)+L_{q3} (\kperp)\right]  \label{lq2}  \quad \textrm{with}\\
 L_{q1} (\kperp) &= -\int\frac{\dd[2]\rperp}{2\pi}e^{-i\kperp\cdot \rperp}S(\rperp)  \biggl(\ln \frac{\kperp^2 \rperp^2}{c_0^2}\biggr)^2 , \label{q2log}\\%\notag \\
 L_{q2} (\kperp) &= (2\pi) F(\kperp) \int \dd[2]\lperp F(\kperp-\lperp)\biggl(\ln\frac{\kperp^2}{\lperp^2}\biggr)^2 , \\% \notag \\
 L_{q3} (\kperp) &= -4 \int \dd[2]\lperp \dd[2]\lpperp F(\kperp - \lpperp)F(\kperp-\lperp)\frac{\lpperp \cdot \lperp}{\lpperp^2 \lperp^2}\ln\frac{\kperp^2}{\lperp^2} .% \notag
\end{align}
In deriving the above expression, we have used the fact that the impact parameter integration simply gives $\targetarea$, which is the area of the target nucleus. (see Appendix for detailed derivations.)

\begin{figure}[tbp]
 \centering
 \includegraphics{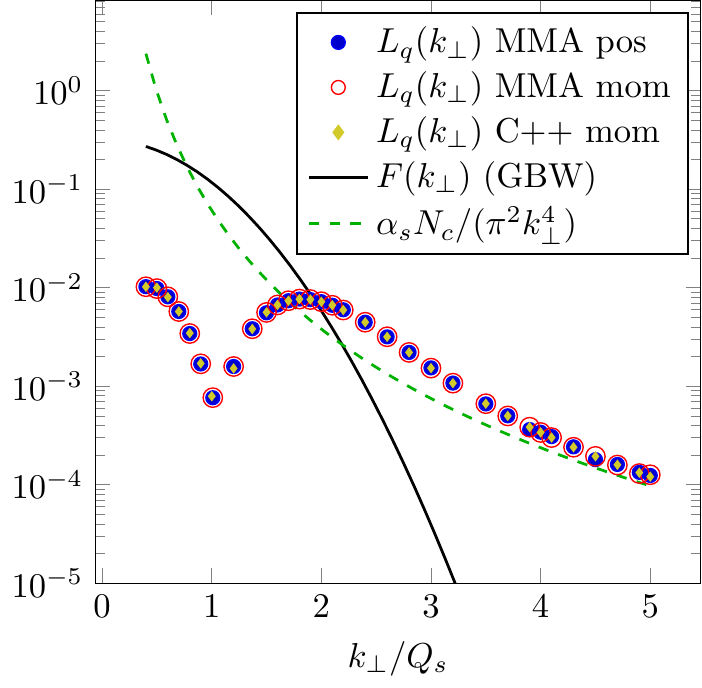}
 \includegraphics{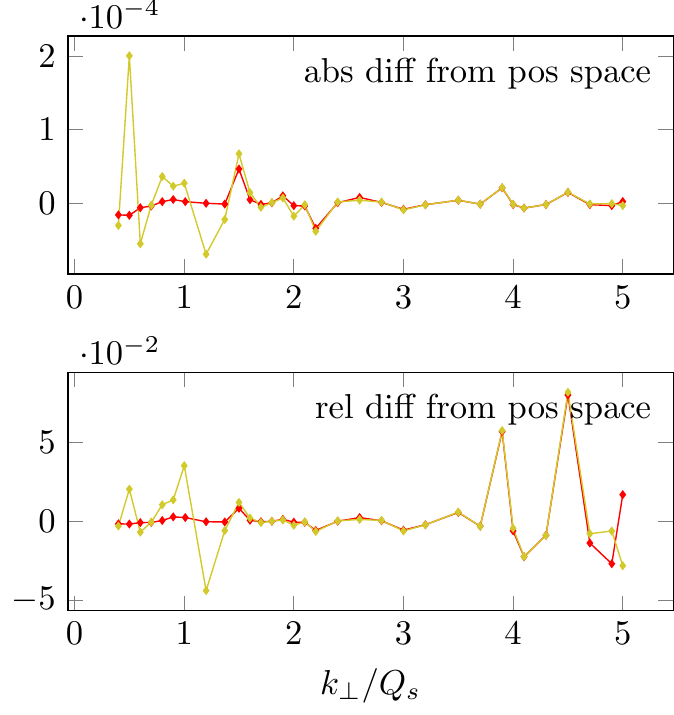}
 \caption[*]{The comparison between $\frac{\satscale^2}{S_\perp}L_q (\kperp)$ and $\satscale^2F(\kperp)$ with $S_\perp$ factored out. (One can also simply set $\satscale=1\, \textrm{GeV}$.) Here we have employed three different numerical methods to evaluate $L_q (\kperp)$. The blue dots indicate the direct numerical evaluation of $L_q(\kperp)$ as in Eq.~(\ref{lq1}) with Mathematica, while the red circles represent the evaluation of Eq.~(\ref{lq2}). The golden diamonds correspond to the numerical results obtain from Eq.~(\ref{lq2}) by using our SOLO code programmed with $C++$. The asymptotic $\kperp$ behaviour of $L_q (\kperp)$ is indicated by the green dashed line. The numerical uncertainties are very small as shown in the right plot.}
\label{co}
\end{figure}

In fact, one can further evaluate $L_q (\kperp)$ analytically in the GBW model by assuming 
\begin{equation}\label{eq:gbw}
S(\Rperp)=\exp \biggl(-\frac{\satscale^2 R_\perp^2}{4}\biggr), \quad \implies \quad F(\kperp)=\frac{1}{\pi \satscale^2} \exp\biggl(-\frac{\kperp^2}{\satscale^2}\biggr)\, ,
\end{equation}
and find 
\begin{align}
L_{q1} (\kperp) &= \frac{-2}{\satscale^2}\left\{L^{(2,0)}\biggl[-1, -\frac{\kperp^2}{\satscale^2}\biggr] -2 \ln \frac{k^2_\perp e^{\gamma_E}}{\satscale^2}L^{(1,0)}\biggl[-1, -\frac{\kperp^2}{\satscale^2}\biggr]\right. \notag \\
                &\hspace{6cm}\left.+\biggl[\biggl(\ln \frac{k^2_\perp e^{\gamma_E}}{\satscale^2}\biggr)^2+\frac{\pi^2}{6}\biggr] \exp\left(-\frac{\kperp^2}{\satscale^2}\right)\right\}, \notag\\
L_{q2} (\kperp) &= 4\pi^2 F(\kperp)F(\kperp) \int_0^\infty \dd\lperp \lperp \exp \left(-\frac{\lperp^2}{\satscale^2}\right) \besselIzero\left(\frac{2\lperp \kperp}{\satscale^2}\right) \biggl(\ln\frac{\kperp^2}{\lperp^2}\biggr)^2 ,\notag \\
L_{q3} (\kperp) &= -\frac{8\pi}{\kperp}F(\kperp) \left[1- \exp \left(-\frac{\kperp^2}{\satscale^2}\right) \right]  \int _0^\infty \dd\lperp  I_1\left(\frac{2\lperp \kperp}{\satscale^2}\right)\exp \left(-\frac{\lperp^2}{\satscale^2}\right)  \ln\frac{\kperp^2}{\lperp^2} .
 \end{align}
For $L_{q2} (\kperp)$ and $L_{q3} (\kperp)$, in principle, one can also perform the $\dd \lperp$ integration and obtain analytical final results in terms of derivatives of hyper-geometrical functions. Asymptotically, $L_{q1} (\kperp)$ and $ L_{q3} (\kperp)$ give $\frac{8\satscale^2}{\kperp^4}$ and $-\frac{4\satscale^2}{\kperp^4}$ in the large $\kperp$ limit, respectively,
while $ L_{q2} (\kperp)$ is exponentially suppressed. The most interesting observation is that $\eval{L_q(\kperp)}_{\kperp\to \infty}=\frac{\alphas \Nc \targetarea}{4\pi^2} \frac{ 4\satscale^2}{\kperp^4}$. 
Comparing to the NLO quark channel hard factor, which involves 
\begin{equation}
 \frac{\alphas \Nc}{4\pi^2} \int_{\tau/z}^1 \dd\xi\, xq(x) \frac{(1+\xi^2)^2}{(1-\xi)_+} \frac{\satscale^2}{\kperp^4}\sim -\frac{\alphas \Nc}{4\pi^2} \xprojectile q(\xprojectile)\frac{61}{12} \frac{\satscale^2}{\kperp^4}\,
\end{equation}
in the large $\kperp$ limit, it is conceivable that this term is sufficient to largely cancel the large and negative NLO cross section found earlier. In order to understand the importance of the additional contribution $\frac{\dd[3]\sigma_{L_q}}{\dd{\rapidity}\dd[2]\pperp}$ as in Eq.~\eqref{sigl}, it is illuminating to compare it with the leading order cross section in the quark channel which can be written as 
\begin{equation}
\frac{\dd[3] \sigma^q_\text{LO}}{\dd\rapidity\dd[2]\pperp} =\int_\tau^1\frac{\dd z}{z^2} \sum_f \xprojectile q_f(\xprojectile) D_{h/q} (z) \targetarea F(\kperp), \quad \textrm{with} \quad F(\kperp) \equiv \int \frac{\dd[2] \rperp  }{(2\pi)^2} e^{-i\kperp \cdot \rperp} S(\rperp ),
\end{equation}
since the LO cross section can provide an order of magnitude estimate of the total cross section. It is then straightforward to see that one just needs to compare $L_q(\kperp)$ (after factorizing out $S_\perp$) with $F(\kperp)$.
As shown in Fig~\ref{co}, $L(\kperp)$ is small compared to $F(\kperp)$ in the low $\kperp$ region, therefore the additional contribution is negligible when $\kperp < \satscale$. One the other hand, $L_q(\kperp)$ falls slowly with $\kperp$ and becomes important when $\kperp > 2 \satscale$. 

\section{The gluon channel contribution}
For the gluon channel, we can use the same procedure to take into account the kinematical constraint. As discussed above, all the hard factors were computed in Refs.~\cite{Chirilli:2011km, Chirilli:2012jd} except for the rapidity divergent piece.
For a large nucleus target and in the large $\Nc$ limit, we can write the rapidity divergent term as  
\begin{multline}
\frac{\alphas\Nc}{\pi^2}\int_{0}^{1}\frac{\dd\xi}{1-\xi }\int \frac{%
\dd[2]\xperp\dd[2]\yperp\dd[2]\bperp}{(2\pi )^{2}}e^{-i\kperp\cdot (\xperp-\yperp)}  \bigl[ -S(\xperp,\yperp)+S(\xperp,\bperp)S(\bperp,\yperp)\bigr]S(x_{\perp },y_{\perp
}) \\
\times\left\{\frac{\left[1-\besselJzero\left(\uperp \Delta \right)\right]^2}{\uperp^{2}} +\frac{\left[1-\besselJzero\left(\upperp \Delta \right)\right]^2}{\upperp^2}-\frac{2\uperp\cdot\upperp}{\uperp^2\upperp^2}\left[1-\besselJzero\left(\uperp \Delta \right)\right]\left[1-\besselJzero\left(\upperp \Delta \right)\right]\right\}, 
\end{multline}
where $\uperp \defn \xperp - \bperp$ and $\upperp\defn \yperp - \bperp$. We have put in the kinematical constraint in the same fashion as we did for the quark channel. Again, after subtracting the small-$x$ logarithm $\ln\frac{1}{x_g}$ through the BK evolution equation for the dipole scattering amplitude in the adjoint representation appearing in the gluon channel, we can obtain the remaining additional hard correction due to the kinematical constraint
\begin{multline}
L_g (\kperp) = \frac{\alpha_s N_c}{\pi^2}\int \frac{\dd[2] \xperp \dd[2]\yperp \dd[2] \bperp}{(2\pi)^2} e^{-i\kperp \cdot (\xperp-\yperp)} \left[ -S(\xperp, \yperp)+S(x_{\perp },b_{\perp})S(b_{\perp },y_{\perp
})\right]S(x_{\perp },y_{\perp
}) \\
\times  \left[\frac{1}{\uperp^2} \ln\frac{\kperp^2 \uperp^2}{c_0^2}+\frac{1}{\upperp^2} \ln\frac{\kperp^2 \upperp^2}{c_0^2} -\frac{2\uperp \cdot u^\prime_\perp}{\uperp^2u^{\prime2}_\perp} \ln\frac{\kperp^2 \abs{\uperp}\abs{\upperp}}{c_0^2}\right]. \label{lad2}
\end{multline}
At the end of the day, we can find the following additional contributions from the gluon channel
\begin{equation}
\frac{\dd[3] \sigma_{L_g}}{\dd\rapidity \dd[2]\pperp} =\int_\tau^1\frac{\dd z}{z^2} \xprojectile  g(\xprojectile) D_{h/g} (z) L_g(\kperp), 
\end{equation}
with $L_g (\kperp)= \frac{\alpha_s N_c}{4\pi^2} S_\perp \left[L_{g1} (\kperp)+L_{g2} (\kperp)+L_{g3} (\kperp)\right]$ and
\begin{align}
 L_{g1} (\kperp)
  &= -2\int\frac{\dd[2]\rperp}{2\pi}e^{-i\kperp\cdot \rperp}S(\rperp)S(\rperp )  \biggl(\ln \frac{\kperp^2 \rperp^2}{c_0^2}\biggr)^2 , \\
 L_{g2} (\kperp)
  &= (4\pi)  \int \dd[2]\lperp  \dd[2]\lpperp F(\kperp-\lperp-\lpperp)F(\kperp-\lpperp)F(\lpperp)\biggl(\ln\frac{\kperp^2}{\lperp^2}\biggr)^2 , \\
 L_{g3} (\kperp)
  &= - 8\int \dd[2]\lperp \dd[2]\lpperp \dd[2]\lppperp F(\lppperp)F(\kperp -\lpperp - \lppperp)F(\kperp-\lperp-\lppperp)\frac{\lpperp\cdot \lperp}{\lpperp^2 \lperp^2}\ln\frac{\kperp^2}{\lperp^2}
\end{align}
In the GBW model, using Eq.~\eqref{eq:gbw}, these expressions can be simplified to 
\begin{align}
 L_{g1} (\kperp)
  &=
  \begin{multlined}[t][\textwidth-10em]
   -\frac{2}{\satscale^2}\Biggl\{L^{(2,0)}\biggl[-1, -\frac{\kperp^2}{2\satscale^2}\biggr] -2 \ln \frac{k^2_\perp e^{\gamma_E}}{2\satscale^2}L^{(1,0)}\biggl[-1, -\frac{\kperp^2}{2\satscale^2}\biggr] \\
    +\biggl[\biggl(\ln \frac{k^2_\perp e^{\gamma_E}}{2\satscale^2}\biggr)^2+\frac{\pi^2}{6}\biggr] \exp\biggl(-\frac{\kperp^2}{2\satscale^2}\biggr)\Biggr\},
  \end{multlined}\\
 L_{g2} (\kperp)
  &=\frac{16\pi}{\satscale^4}F(\kperp)  \int _0^\infty \dd\lperp \lperp  \int_0^\infty \dd\qperp \qperp \besselIzero\biggl(\frac{2\kperp \qperp}{\satscale^2}\biggr) \besselIzero\biggl(\frac{2\lperp \qperp}{\satscale^2}\biggr) e^{-\frac{\lperp^2+3\qperp^2}{\satscale^2}}
  \biggl(\ln\frac{\kperp^2}{\lperp^2}\biggr)^2 , \\
 L_{g3} (\kperp)
  &= -\frac{32\pi}{\satscale^2}F(\kperp)  \int _0^\infty \dd\lperp  \int_0^\infty \dd\qperp \bigl(1- e^{-\qperp^2/\satscale^2} \bigr) \besselIzero\biggl(\frac{2\kperp \qperp}{\satscale^2}\biggr) I_1\biggl(\frac{2\lperp \qperp}{\satscale^2}\biggr) e^{-\frac{\lperp^2+2\qperp^2}{\satscale^2}}
  \ln\frac{\kperp^2}{\lperp^2} . 
\end{align}
It is straightforward to find that the leading power behaviour of $L_g (\kperp)$ is $\frac{\alpha_s N_c}{4\pi^2} S_\perp\frac{ 8\satscale^2}{\kperp^4}$ with $\satscale$ defined as the quark saturation momentum.

\section{Numerical results and physical discussions}

\subsection{Numerical setup}

The calculation of single inclusive hadron production in $\pA$ collisions up to NLO level with the full running coupling has been recently implemented as the computer program Saturation physics at One Loop Order, or SOLO~\cite{Stasto:2013cha}. Initial results from the program showed that the full NLO $\pA\to hX$ cross section agrees fairly well with the forward RHIC data. The same paper also provided numerical results at the energy scale of the LHC at extreme forward rapidities $\rapidity \gtrsim 5$, since at the time no LHC experiment had published any results.

Since then, ALICE, CMS, and ATLAS have released their experimental data of the $\pA\to hX$ differential cross section, all of which are at roughly central rapidity, $-2 \lesssim \rapidity \lesssim 2$. Unfortunately, the initial version of SOLO gives results with very large uncertainties for LHC central rapidity collisions, for basically the following reasons: first, several of the terms involve oscillatory factors of the form $J_0\bigl(\frac{\pperp \rperp}{z}\bigr)$ or $J_0\bigl(\frac{\pperp \rperp}{z\xi}\bigr)$, which are integrated over $\rperp$.
At the LHC, the physical scenario where a high $\xprojectile$ parton from the proton projectile radiates a soft (small $\xi$) gluon (or quark), which then fragments into the produced hadron, becomes a much more significant contribution than at RHIC.
Specifically, $\xi$ or $z$ can be as small as $\tau = \frac{\pperp}{\sqsnn} e^\rapidity$, which decreases from $0.04$ at BRAHMS with $\rapidity = 2.2$, to $0.0002$ at ALICE or ATLAS with $\rapidity = 0$ for $\pperp = \SI{1}{GeV}$.
Such small values of $z$ and $\xi$ produce rapid oscillations in the integrand, which most numerical integration algorithms are notoriously bad at handling.
Although there are specialized algorithms available, they are prohibitively difficult when the integrals have 4, 6, or even 8 dimensions, as is the case with most terms in the cross section. 

Other terms involve factors like $F(\kperp - \lperp)$ which are integrated over $\lperp$, or similar factors with other forms of the argument. These functions have their peaks where the argument is zero: for example, $\lperp = \kperp$ in the first case. The numerical integration algorithms used by SOLO are most effective when the function being integrated has its peak where the integration variable $\lperp = 0$, which is approximately satisfied at RHIC (the largest contributions come from $\kperp\sim \SI{10}{GeV}$), but not at the LHC. Under LHC kinematical conditions, we have to accommodate much larger values of $\kperp$ while keeping the numerical uncertainties under control.

Through a series of transformations and various tricks, together with a tremendous amount of effort, we have converted the formulas originally used by SOLO~\cite{Zaslavsky:2014asa} to improved versions with vastly smaller numerical uncertainties. At this moment, we believe that all the numerical uncertainties are well under control even at the middle rapidity region of the LHC. Appendix B explains some of the key points of these transformations, but we will reserve the full details of the numerical implementation for a separate document to be released later.

% talking about dipole cross section parametrizations

We have computed results for two different parametrizations of the dipole scattering amplitude $S$ (or its momentum space expression, $F$). First, the GBW model, defined by
\begin{equation}
 S_{\text{GBW}}(\rperp)=\exp \biggl(-\frac{\satscale^2 \rperp^2}{4}\biggr)
\end{equation}
with
\begin{equation}
 \satscale^2(\xtarget)=\centrality\massnumber^{1/3}Q_0^2\biggl(\frac{x_0}{\xtarget}\biggr)^{\lambda}
\end{equation}
where $A$ represents the number of nucleons in the target nucleus and $c = \sqrt{1 - b^2/R^2}$, where $b$ is the impact parameter and $R$ is the nuclear radius, accounts for the centrality of the collision. We use the same parameters as in previous SOLO results~\cite{Stasto:2013cha,Stasto:2014sea,Zaslavsky:2014asa}: $c = 0.56$ to represent minimum bias collisions, and the original GBW fit parameters from Ref.~\cite{GolecBiernat:1998js}, $x_0=0.000304$, $\lambda=0.288$ and $Q_0^2=\SI{1}{GeV^2}$, which are based on HERA data. The GBW model is often used in phenomenological calculations due to its simple analytical form.

In addition, we also show results computed using the numerical solution of the BK equation~\cite{Balitsky:1995ub, Kovchegov:1999yj} with the running coupling correction~\cite{Balitsky:2006wa, Balitsky:2008zza, Kovchegov:2006wf, Kovchegov:2006vj}, setting the QCD running coupling scale in the rcBK equation to $\Lambda=\SI{0.1}{GeV}$.
This solution has been shown~\cite{GolecBiernat:2001if, Albacete:2007yr, Berger:2010sh, vanHameren:2014lna} to be very useful in phenomenology, especially at high transverse momenta.
Previous studies have also considered the McLerran-Venugopalan (MV) model~\cite{MV}, an analytical expression which gives a high-$\kperp$ power tail similar to the BK solution, but we have omitted those results from the present analysis because the GBW and rcBK models are sufficient to show the interesting features of the results.

% discussion of figures

Results computed from SOLO are for the quantity $\frac{1}{\targetarea}\frac{\dd[3] \sigma^{\pA\to \pi X}}{\dd\rapidity\dd[2] \pperp}$, which can be trivially converted into the differential yield for production of a single pion species in the center-of-mass frame,
\begin{equation}
 \frac{\dd[3] N^{\pA\to \pi X}}{\dd\rapidity\dd[2] \pperp}
 =
 \frac{\targetarea}{\sigma_\text{inel}} \frac{1}{\targetarea}\frac{\dd[3] \sigma^{\pA\to \pi X}}{\dd\rapidity\dd[2] \pperp}
\end{equation}
where $\pi$ is a single species of pion: $\pi^+$, $\pi^0$, or $\pi^-$.
However, the experiments measure
\begin{equation}
 \frac{1}{2\pi\pperp}\frac{\dd[2] N^{\pA\to h X}}{\dd\pseudorapidity\dd\pperp}
 =
 \frac{1}{2\pi\pperp \sigma_\text{inel}}\frac{\dd[2] \sigma^{\pA\to h X}}{\dd\pseudorapidity\dd\pperp}
\end{equation}
where $h$ may include several different hadron species, depending on the detector, and $\sigma_\text{inel}$ is the \emph{total} inelastic cross section.
We neglect the difference between rapidity in the lab frame and pseudorapidity $\pseudorapidity$.
Accordingly, we multiply the output from SOLO by a factor $\frac{\sigma^h}{\sigma^\pi}\targetarea/\sigma_\text{inel}$ to make it compatible with the experimental measurements.
%(Note that this is not a ``$K$-factor'' in the traditional sense, which is purely phenomenological and has no theoretical justification.)
\begin{itemize}
 \item BRAHMS measures negatively charged hadrons. We use $\frac{\sigma^h}{\sigma^\pi} = 1.3$ to account for the yields from kaons and other hadrons. We also set $\targetarea \approx \pi(\SI{7.5}{fm})^2 = \SI{1770}{mb}$, and use $\sigma_\text{inel} = \SI{2400}{mb}$~\cite{Debbe:2004ci}.
 \item STAR measures only neutral pions, so $\frac{\sigma^h}{\sigma^\pi} = 1$, and $\sigma_\text{inel} = \SI{2210}{mb}$~\cite{Adams:2006uz}, with the same $\targetarea$ as BRAHMS.
 \item ALICE and ATLAS measure all charged pions, kaons, and protons. We use a result from CMS~\cite{Chatrchyan:2013eya} that the kaon and proton yields are 13\% and 6\%, respectively, of the pion yield, giving $\frac{\sigma^h}{\sigma^\pi} \approx 2.4$, and $\sigma_\text{inel} = \SI{2100}{mb}$ from LHCb~\cite{LHCb:2012aka}. For lead nuclei, $\targetarea \approx \targetarea^\text{Au}\times(208/197)^{2/3} = \SI{1830}{mb}$.
\end{itemize}

\begin{figure}
 \centering
 \includegraphics{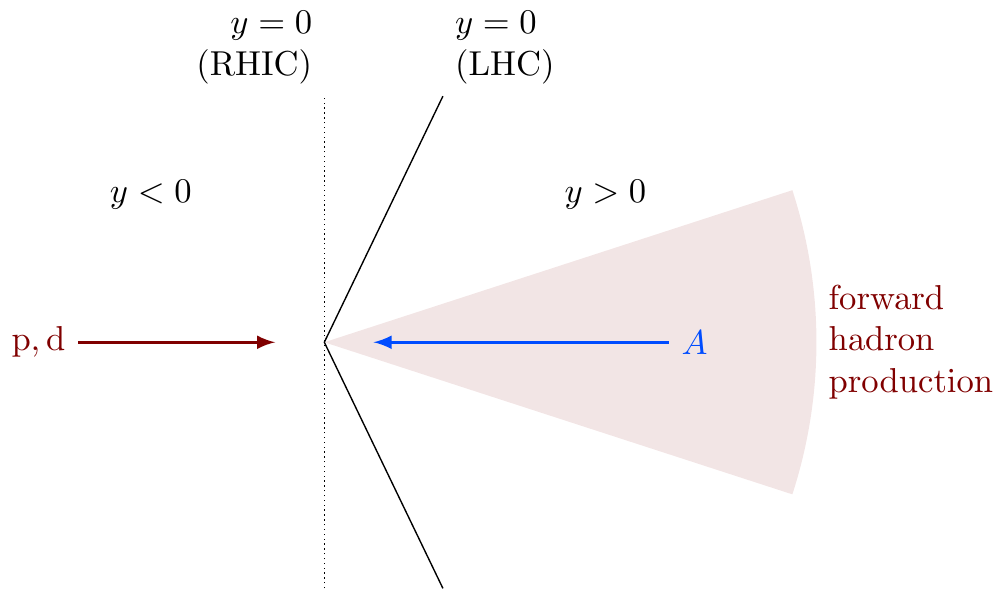}
%  \tikzsetnextfilename{beamsetup}
%  \begin{tikzpicture}
%   \fill[red!50!black,fill opacity=0.1] (0,0) -- (18:5) arc[radius=5cm,start angle=18,end angle=-18] node[pos=0.5,right,align=left,opacity=1] {forward\\hadron\\production} -- cycle;
%   \tikzset{z={(1cm,0)},y={(0,0)},x={(0,1cm)}}
%   \draw[densely dotted] (detector rectangle cs:0,180,2.5,2) -- (0,0) -- (detector rectangle cs:0,0,2.5,2) node[above left,align=right] {$\rapidity = 0$\\(RHIC)};
%   \draw[thin] (detector rectangle cs:0.465,180,2.5,2) -- (0,0) -- (detector rectangle cs:0.465,0,2.5,2) node[above right,align=left] {$\rapidity = 0$\\(LHC)};
%   \node (lab rapidity top 1) at (detector rectangle cs:1.4,0,1.5,2) {$\rapidity > 0$};
%   \node (lab rapidity top -1) at (detector rectangle cs:-1,0,1.5,2) {$\rapidity < 0$};
%   \draw[red!50!black,thick,latex-] (0,0,-0.5) -- +(0,0,-2) node[left] {$\mathrm{p},\mathrm{d}$};
%   \draw[blue!70!cyan,thick,latex-] (0,0,0.5) -- +(0,0,3) node[right] {$A$};
%  \end{tikzpicture}
 \caption[*]{The orientation of rapidities used by SOLO and throughout this paper. Positive (or forward) rapidity is always in the direction of the proton (or deuteron, for RHIC) beam. Some results published by ALICE~\cite{ALICE:2012mj} and ATLAS~\cite{AlexanderMilovonbehalfoftheATLAS:2014rta} use the opposite orientation. In this paper we always use $\rapidity$ to represent the rapidity in the center-of-mass frame.}
 \label{fig:beamsetup}
\end{figure}

As far as the definition of rapidities is concerned, deuteron beams have positive rapidity and gold nuclei beams have negative rapidity at RHIC. The energy of both beams per nucleon is $\sqsnn=\SI{200}{GeV}$. Therefore, the center-of-mass frame is the same as the lab frame. The specification of rapidities in the SOLO package follows the above setup: positive rapidity always refers to the deuteron-going direction (or proton-going, at the LHC), as shown in Fig.~\ref{fig:beamsetup}. On the other hand, some of the ATLAS and ALICE $\pPb$ data measured at $\sqsnn=\SI{5.02}{TeV}$ are presented in the opposite rapidity configuration, with the proton beams having negative rapidity. In order to compare our results with those data without confusion, we have flipped the sign of rapidity in the experimental results and labeled our plots with the rapidity $\rapidity$ in the center-of-mass frame in the SOLO convention. Throughout this paper, forward rapidity, $\rapidity > 0$, always means the rapidity region along the proton (or deuteron) beam direction in the center-of-mass frame. 

\subsection{Discussion of numerical results}

\begin{figure}[tbp]
 \centering
 \includegraphics{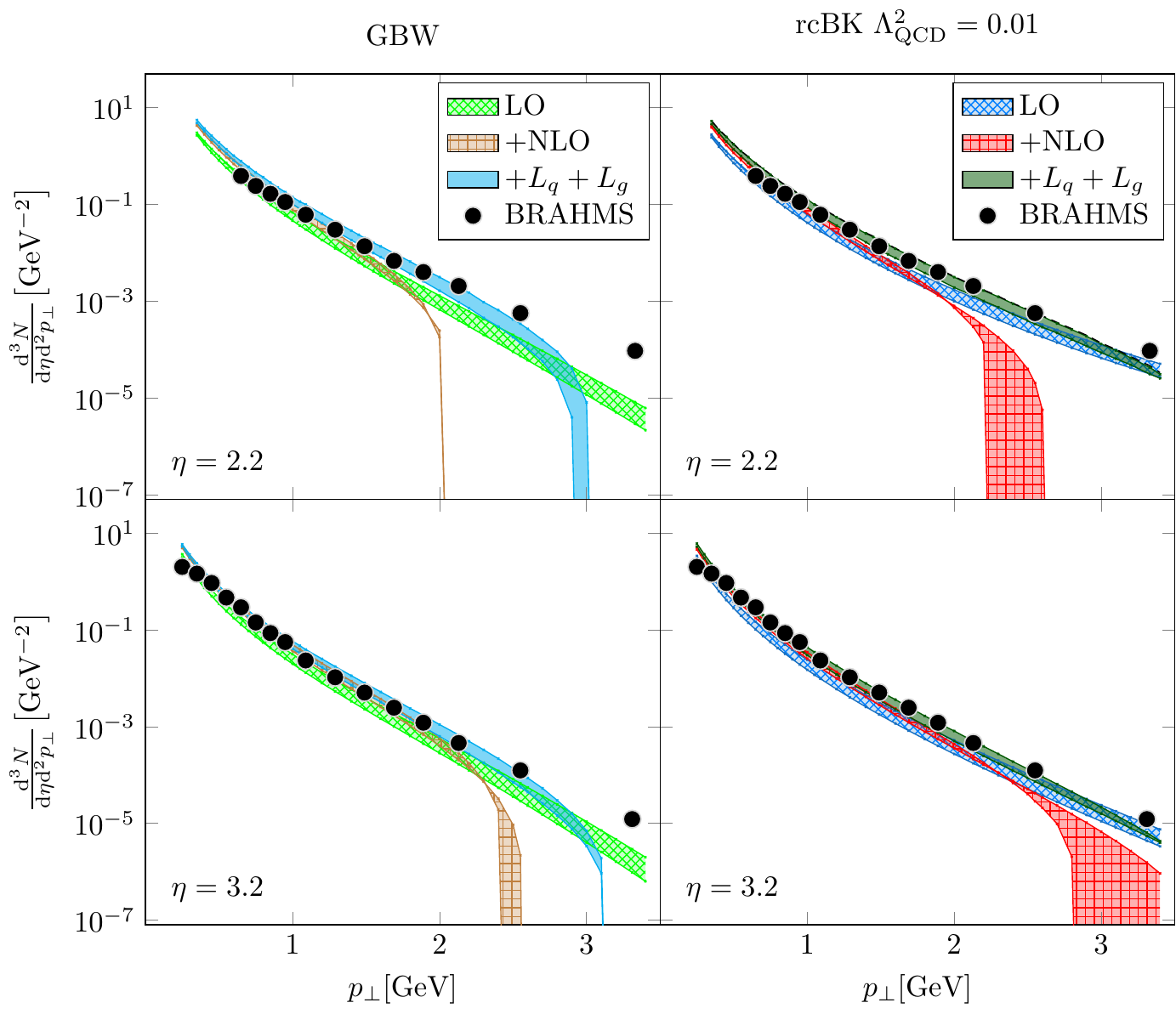}
 \caption[*]{Comparisons of BRAHMS data~\cite{Arsene:2004ux} with the center-of-mass energy of $\sqsnn=\SI{200}{GeV}$ per nucleon at rapidity $\rapidity=2.2, 3.2$ with our results. As illustrated above, the crosshatch fill shows LO results, the grid fill indicates LO+NLO results, and the solid fill corresponds to our new results which include the NLO corrections from $L_q$ and $L_g$ due to the kinematical constraint. The error band is obtained by changing $\mu^2$ from $\SI{10}{GeV^2}$ to $\SI{50}{GeV^2}$.}
 \label{g}
\end{figure}

\begin{figure}
 \centering
 \includegraphics{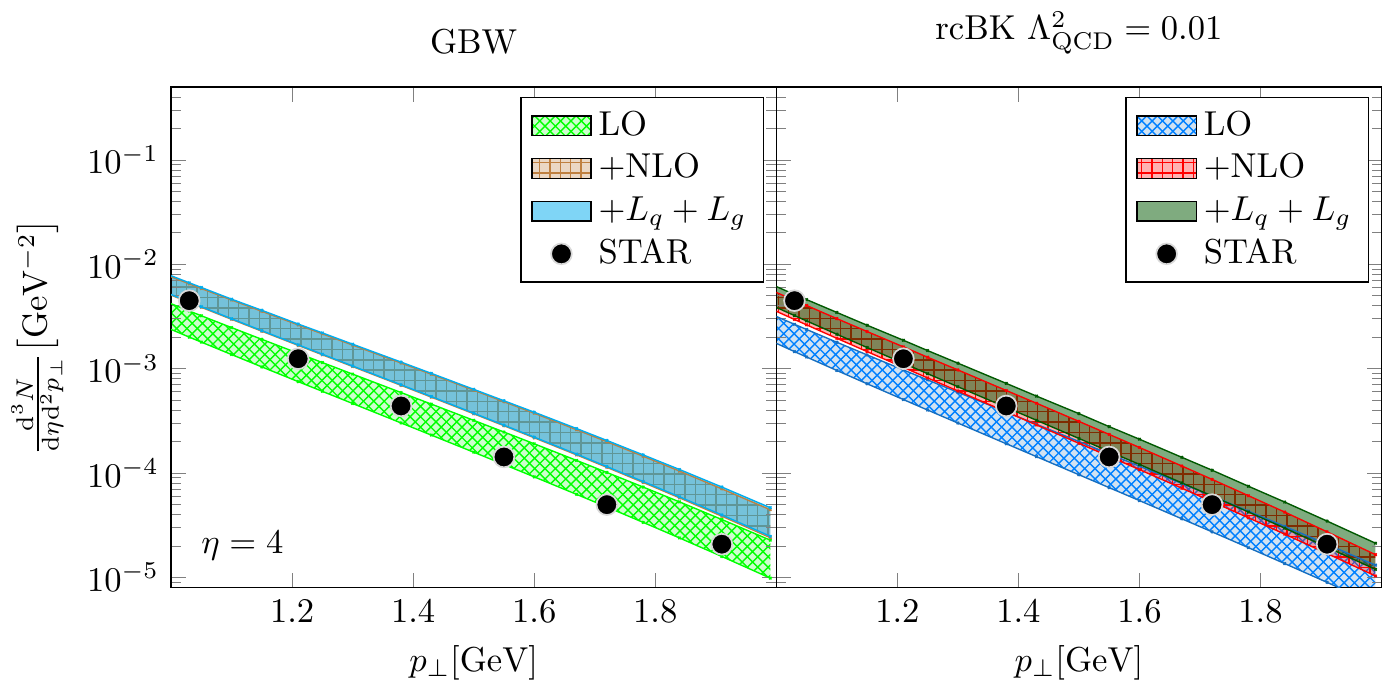}
%  \tikzsetnextfilename{star_combined2x1}
%  \begin{tikzpicture}
%   \begin{groupplot}[width=0.47\linewidth,xtick placement tolerance=-0.05pt,result axis,ymode=log,ymin=8e-6,ymax=5e-1,xmin=1,xmax=2,group style={group size={2 by 1},horizontal sep=0pt,vertical sep=0pt,x descriptions at=edge bottom,y descriptions at=edge left}]
%   \nextgroupplot[title={GBW}]
%    \resultplot{GBW LO,star xsec LO}{starGBW}
%    \resultplot{GBW NLO,star xsec NLO}{starGBW}
%    \resultplot{GBW extra,star xsec Lqg}{starGBW}
%    \addplot[data plot,star data] table {\stardAu};
%    \legend{$\text{LO}$,$+\text{NLO}$,$+L_q+L_g$,STAR}
%    \node[anchor=south west] at (axis description cs:0.03,0.03) {$\eta = 4$};
%   \nextgroupplot[title={rcBK $\Lambda_{\text{QCD}}^2=0.01$}]
%    \resultplot{rcBK LO,star xsec LO}{starrcBK}
%    \resultplot{rcBK NLO,star xsec NLO}{starrcBK}
%    \resultplot{rcBK extra,star xsec Lqg}{starrcBK}
%    \addplot[data plot,star data] table {\stardAu};
%    \legend{$\text{LO}$,$+\text{NLO}$,$+L_q+L_g$,STAR}
%   \end{groupplot}
%  \end{tikzpicture}
 \caption[*]{Comparison of STAR data~\cite{Adams:2003qm} with $\sqsnn=\SI{200}{GeV}$ at $\rapidity=4$ with results from SOLO for the GBW and rcBK models. The color scheme is the same as in figure~\ref{g}, and again, the error band comes from $\mu^2 = \SI{10}{GeV^2}$ and $\SI{50}{GeV^2}$. We do not see the negative total cross section because the cutoff momentum above which the cross section becomes negative is larger than the $\pperp$ of the available data, and in fact larger than the kinematic limit $\sqsnn e^{-\rapidity}$.}
 \label{fig:star}
\end{figure}

\begin{figure}
 \centering
 \includegraphics{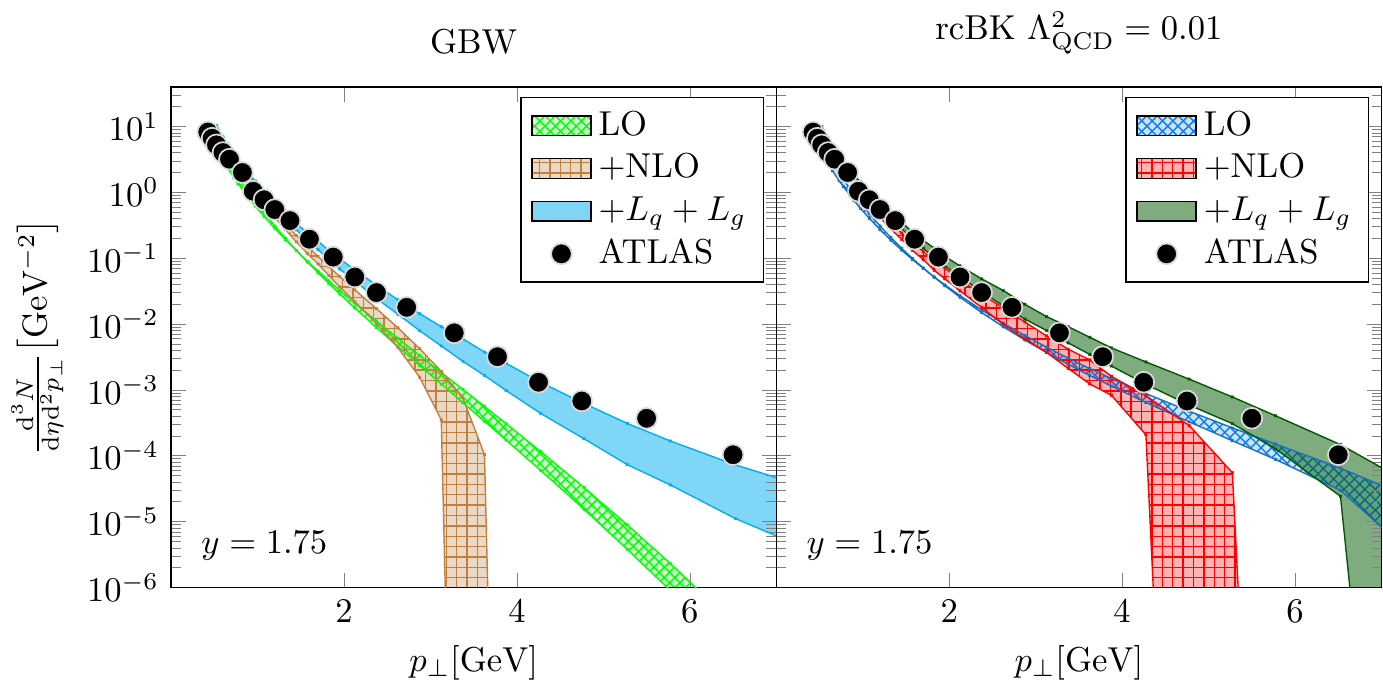}
%  \tikzsetnextfilename{atlas_combined_hiY_2x1}
%  \begin{tikzpicture}
%   \begin{groupplot}[width=0.47\linewidth,xtick placement tolerance=-0.05pt,result axis,ymode=log,ymin=1e-6,ymax=4e1,xmin=0,xmax=7,group style={group size={2 by 1},horizontal sep=0pt,vertical sep=0pt,x descriptions at=edge bottom,y descriptions at=edge left}]
%   \nextgroupplot[title={GBW}]
%    \resultplot{GBW LO,atlas xsec LO}{atlasGBWB}
%    \resultplot{GBW NLO,atlas xsec NLO}{atlasGBWB}
%    \resultplot{GBW extra,atlas xsec Lqg}{atlasGBWB}
%    \addplot[data plot,atlas y175 data] table {\atlaspPb};
%    \legend{$\text{LO}$,$+\text{NLO}$,$+L_q+L_g$,ATLAS}
%    \node[anchor=south west] at (axis description cs:0.03,0.03) {$\rapidity = 1.75$};
%   \nextgroupplot[title={rcBK $\Lambda_{\text{QCD}}^2=0.01$}]
%    \resultplot{rcBK LO,atlas xsec LO}{atlasrcBKB}
%    \resultplot{rcBK NLO,atlas xsec NLO}{atlasrcBKB}
%    \resultplot{rcBK extra,atlas xsec Lqg}{atlasrcBKB}
%    \addplot[data plot,atlas y175 data] table {\atlaspPb};
%    \legend{$\text{LO}$,$+\text{NLO}$,$+L_q+L_g$,ATLAS}
%    \node[anchor=south west] at (axis description cs:0.03,0.03) {$\rapidity = 1.75$};
%   \end{groupplot}
%  \end{tikzpicture}
 \caption[*]{Comparison of ATLAS forward-rapidity data~\cite{AlexanderMilovonbehalfoftheATLAS:2014rta} with the center-of-mass energy of $\sqsnn=\SI{5.02}{TeV}$ at $\rapidity = 1.75$ with SOLO results for the GBW and rcBK models. Again, the color scheme is the same as in figure~\ref{g}. Here the error band shows plots for $\mu^2 = \SI{10}{GeV^2}$ and $\mu^2 = \SI{100}{GeV^2}$. Since the numerical data for these measurements are not published, we have extracted the ATLAS points from Fig.~6 of Ref.~\cite{AlexanderMilovonbehalfoftheATLAS:2014rta}. The extraction procedure introduces uncertainties comparable to the size of the points.}
 \label{fig:atlas_forward}
\end{figure}

\begin{figure}
 \centering
 \includegraphics{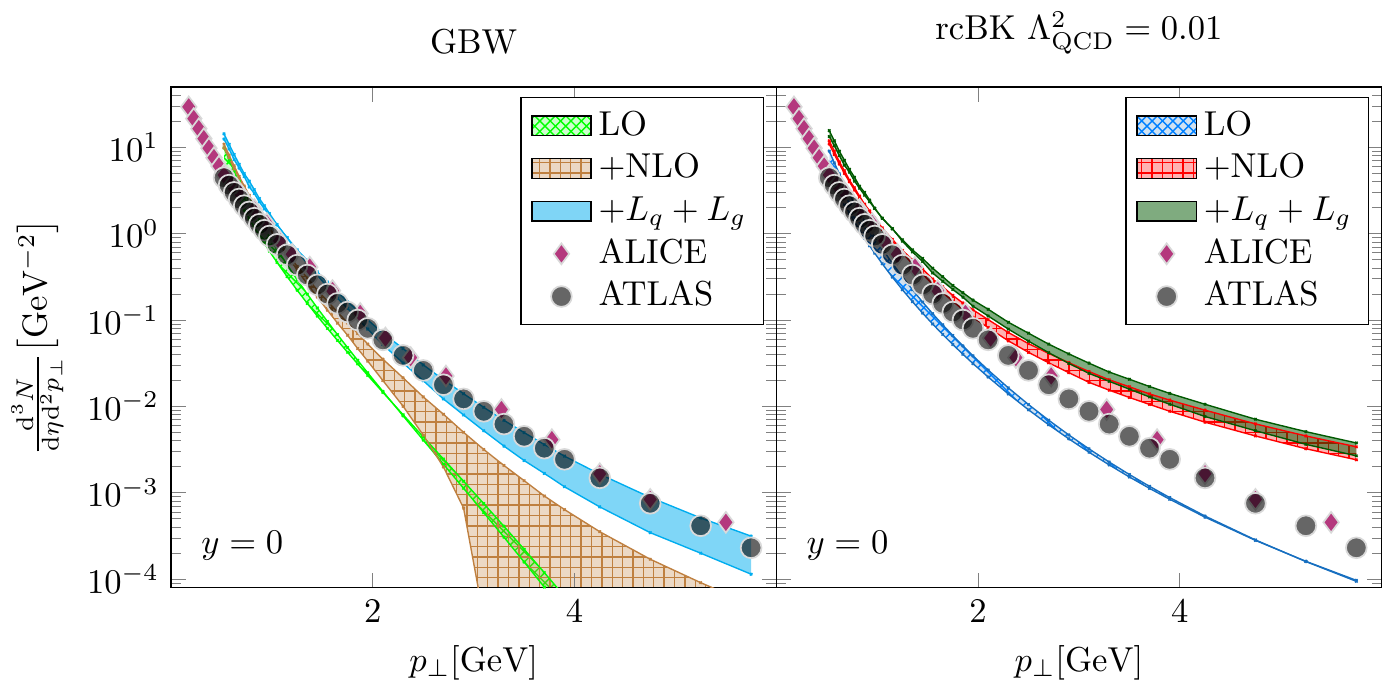}
 \caption[*]{Comparison of mid-rapidity data at $\sqsnn=\SI{5.02}{TeV}$ at $\rapidity=0$ from ALICE~\cite{ALICE:2012mj} and ATLAS~\cite{AlexanderMilovonbehalfoftheATLAS:2014rta} with SOLO results for the GBW and rcBK models. Both data sets correspond to the same kinematic parameters. Again, the color scheme is the same as in figure~\ref{g}. These results display the breakdown of the dilute-dense factorization approach when the separation between $\xprojectile$ and $\xtarget$ is not sufficiently large. $\frac{\xtarget}{\xprojectile}$ is roughly on the order of $1$.}
 \label{fig:alice_atlas_mid}
\end{figure}

Figures~\ref{g} and~\ref{fig:star} show the differential $\dA\to hX$ yields at forward rapidity for BRAHMS~\cite{Arsene:2004ux} and STAR~\cite{Adams:2003qm}, respectively, along with the corresponding results from SOLO using the new (transformed) formulas. The LO and LO+NLO curves are very similar to earlier results published in Ref.~\cite{Stasto:2013cha}; some slight differences are due to the increased precision of the new formulas. In the meantime, the $L_q$ and $L_g$ corrections are completely negligible in the region where $\pperp \lesssim Q_s$. On the other hand, where $\pperp \gtrsim Q_s$, $L_q$ and $L_g$ start to become important and alleviate the negativity problem in the GBW model, and help us to better describe the data in the high $\pperp$ region. In the rcBK case, we find that the full NLO cross section now becomes completely positive and provides us excellent agreement with all the RHIC data.

In Figure~\ref{fig:atlas_forward}, we show the comparison between the forward ATLAS data at $\rapidity=1.75$ and the numerical results from SOLO. We observe remarkable agreement between the full NLO calculation from the saturation formalism and experimental data up to $\SI{6}{GeV}$. Again, as we have seen earlier, the newly added $L_q$ and $L_g$ corrections help to increase the applicable $\pperp$ window of the saturation formalism from roughly \SIrange[range-phrase=--,range-units=single]{2.5}{3}{GeV} to $\SI{6}{GeV}$. From $\SI{6}{GeV}$ and up, the full NLO cross section still becomes negative, which implies that the saturation formalism does not apply anymore and the collinear factorization should be used. Admittedly, what we have seen is only one piece of a promising clue for the gluon saturation phenomenon. More data in different forward rapidity windows at the LHC would allow us to conduct precise tests of the theoretical calculation, and may eventually provide us the smoking gun proof. 

In Figure~\ref{fig:alice_atlas_mid}, we show the comparison between the ALICE and ATLAS data at $\rapidity=0$ and the numerical results from SOLO. We find that the full NLO results, 
especially the one with the rcBK solution, miss the data. (It seems that the GBW model roughly agrees with the data, but we believe that it is probably just a coincidence.) This indicates that the dilute-dense factorization breaks down at $\rapidity=0$. This is completely expected for the following reason. First, the collinear parton distributions of the proton projectile do not resum small-$x$ logarthms and may have considerable uncertainties in the very low-$x$ region. Most importantly, the dilute-dense factorization derived in Refs.~\cite{Chirilli:2011km, Chirilli:2012jd} assumes that the proton projectile is dilute while the nuclear target is dense. In the forward rapidity region, for example $\rapidity=1.75$, one can estimate that roughly $\xtarget/\xprojectile \sim 10^{-2}$, which indicate that the small-$x$ evolution is more important in the nuclear target than in the proton projectile. Therefore, we can use the integrated parton distributions in the proton projectile and use the small-$x$ evolution for the unintegrated gluon distribution in the nuclear target. On the other hand, in the middle rapidity region, $\xtarget/\xprojectile \sim 1$, we should use the small-$x$ evolution to resum both $\alphas \ln\frac{1}{\xtarget}$ and $\alphas \ln\frac{1}{\xprojectile}$ simultaneously, since they are of the same order. This means that we need to do a complete NLO calculation in the framework of the so-called $\kperp$ factorization with unintegrated parton distributions for both the proton projectile and the nuclear target. Unfortunately, this calculation is very challenging, therefore we shall leave it to future studies. 

Some more discussion and comments are in order, as follows. First of all, as we have shown analytically and numerically in the GBW model, 
the two additional NLO corrections derived in this work extend the applicability of the saturation physics calculation further into the large $\pperp$ region, without significantly modifying the low $\pperp$ results. The numerical results using the rcBK solution also support the same conclusion. In the very large $\pperp$ region where the saturation effect is extremely small, the full NLO cross section may still become negative, but this is already beyond the applicability of the saturation formalism. In this region, it is well-known that the collinear factorization is the relevant framework and provides the best description of the QCD dynamics. 

Second, the comparison between the data and our calculation suggests that the implementation of the kinematical constraint works slightly better for the rcBK approach. As compared to the GBW model, the rcBK solution of the dipole scattering amplitude has the correct perturbative tail. %This is expected since the rcBK approach provides the best description of the data in general. 

Last but not least, as shown in the numerical results, the negative cross section at NLO may still persist when $\pperp$ of the measured hadron is much larger than the saturation momentum~$\satscale$. We recall the results of Ref.~\cite{Stasto:2014sea}: first, that the perturbative QCD calculation from collinear factorization can describe data in the large $\pperp$ region; and also, with exact kinematics (which simply removes the plus function), the perturbative QCD calculation analytically matches the large-$\pperp$ expansion of the results from small-$x$ factorization at next-to-leading order in $\alphas$. Therefore, the nuclear modification factor $\RpA$ measured in $\pA$ collisions at large $\pperp$ ($\pperp\gg \satscale(A)>\satscale(p)$) becomes
\begin{equation}
\RpA \defn \frac{\dd\sigma^{\pA}/\dd \rapidity \dd[2]\pperp}{\Ncoll \dd\sigma^{\pp}/\dd \rapidity \dd[2]\pperp} \simeq \frac{\targetarea^A \satscale^2(A)}{A \targetarea^p \satscale^2(p)} =\frac{xG^A(x)}{A xG^p(x)}\simeq 1.
\end{equation}
Here the number of binary collisions $\Ncoll$ is $\massnumber$ in $\pA$ collisions. ($\Ncoll$ is $\massnumber^{1/3}$ if $\RpA$ is computed from the measured yields.) This has been repeatedly seen in a lot of experimental data, for example PHENIX~\cite{Adler:2003kg}, BRAHMS~\cite{Arsene:2004ux}, STAR~\cite{Adams:2003qm}, ALICE~\cite{ALICE:2012mj}, CMS~\cite{CMS:2013cka} and ATLAS~\cite{AlexanderMilovonbehalfoftheATLAS:2014rta}. As indicated by both the experimental data and our NLO analysis, it seems more and more clear that, at sufficiently high energy and forward rapidity, the saturation effect is dominant in the low $\pperp$ region where $\RpA < 1$. On the other hand, the moderate and large $\pperp$ region, where $\RpA \geq 1$, is described by simple perturbative QCD, with the saturation effects encoded in the subleading power corrections $\mathcal{O}(\frac{\satscale^2}{\pperp^2})$ as shown in e.g. Eqs.~(28) and~(29) of~Ref.\cite{Kharzeev:2003wz}. To search for clear and compelling evidence of gluon saturation in single inclusive hadron production, one should focus on the low-$\pperp$ part of spectra in the forward rapidity region of the proton beam in $pA$ collisions, which is dominated by semi-hard scattering in the vicinity of the saturation scale $\satscale$. Having said that, one should also be aware that $\pperp$ of the measured hadron should be kept sufficiently large (at least~$\SI{0.5}{GeV}$) to avoid nonperturbative QCD effects.

\section{Conclusion}

In this paper, we have investigated the details of  applying the dipole formalism to inclusive hadron production in forward $\pA$ collisions at the energy ranges of RHIC and the LHC. 
In particular, we derived two additional terms by considering the kinematical constraint~\eqref{eq:kinematicalconstraint} in the dipole formalism at next-to-leading order. These two terms were assumed to be negligible in the high energy limit $s\to \infty$ in previous studies. 
In order to do more precise and reliable numerical calculations for phenomenological studies of saturation physics, we have to include these additional terms at finite center-of-mass energy $\sqsnn$. 
From an extensive phenomenological study, we found that these additional terms extend the applicability of the NLO cross section in the small-$x$ saturation physics formalism to higher values of the produced hadron momentum $\pperp$.
Matching to the perturbative collinear factorization will further extend the kinematic coverage of the theoretical predictions for this process~\cite{Stasto:2014sea}.

From the explicit NLO analysis of the single inclusive hadron spectrum in $\pA$ collisions, we argue that the nuclear modification factor $\RpA$ shall approach $1$ at sufficient large transverse momentum, where the collinear factorization calculations apply to both $\pp$ and $\pA$ collisions.
In the low transverse momentum region $\pperp\leq \satscale$, the small-$x$ factorization approach is the appropriate framework to compute inclusive hadron production in $\pA$ collisions, where the gluon saturation effects can be systematically included.
Therefore, this process $\pA\to h+X$ provides a unique opportunity to investigate the interplay between two important QCD dynamical effects (small-$x$ resummation and the collinear factorization calculation) in high energy hadronic reactions. 

In our calculations, by including the kinematical constraint and the newly added NLO correction terms, we are able to improve the results of SOLO and achieve excellent agreement with the forward rapidity RHIC data in $\dAu$ collisions. Furthermore, we have significantly improved the numerical accuracy of the SOLO package, which allows us to compute forward rapidity observables at LHC energy with small uncertainties and obtain remarkable agreement with the forward rapidity ATLAS data at $\sqsnn=\SI{5.02}{TeV}$ in the relatively low-$\pperp$ region.
These results could be additional compelling evidence for the observation of the onset of saturation effects at the LHC.

Our results provide a benchmark framework for the small-$x$ saturation calculation for high energy processes in $\pA$ collisions at the next-to-leading order.
Recent theoretical developments have revolutionized the test of the saturation physics from the qualitative level to the quantitative level.
With more and more experimental data available from the LHC, we will be able to tell whether and when gluon saturation has an effect at extremely small $x$ and large nucleus mass numbers.
The theoretical advances in computing these processes at the next-to-leading order will be crucial to help identify the signature of gluon saturation phenomena.
We expect more developments along the line discussed in this paper.

Phenomenologically, the SOLO program has been developed from a single-purpose program, for specific formulas under RHIC conditions only, to a more general-purpose program that can easily be adapted to different expressions and produce results for both RHIC and the LHC.
During the development process, we have found an efficient method to perform the numeric computations involve.
This will become useful and applicable to other interesting processes. 

\section*{Acknowledgements}
B.X. and D.Z. wish to thank the nuclear theory group at the Lawrence Berkeley National Laboratory for hospitality and support during their visit while this work was in preparation. D.Z. would also like to thank the Penn State Institute for Cyber Science for computational resources essential to the completion of this project. F. Y. is supported by the U.S. Department of Energy, Office of Science, Office of Nuclear Physics, under contract number DE-A C02-05CH11231. We thank T. Altinoluk, N. Armesto, A. Kovner, A. Mueller and A. Sta\'sto for stimulating discussion.

\appendix
\section{The evaluation of some integrals}
First of all, all the Fourier transform formula used in this paper can be derived with the following two identities for $\uperp >0$ together with analytical continuation
\begin{align}
G(\alpha) \defn \int_0^\infty \dd\lperp \lperp^{1+\alpha} \besselJzero(\lperp \uperp) &=\frac{2^{1+\alpha}\Gamma [1+\frac{\alpha}{2}]}{\uperp^{2+\alpha}\Gamma [-\frac{\alpha}{2}]} \\
\int_0^\infty \dd\lperp \lperp \besselJzero(\lperp \uperp) \ln^n \lperp &= \eval{\dv[n]{G(\alpha)}{\alpha}}_{\alpha\to 0}.
\end{align}
Let us now provide some essential details for deriving $L_1(\kperp)$, which comes from the following integral
\begin{align}
\MoveEqLeft
 \begin{multlined}-\frac{\alphas \Nc}{\pi^2}\int \frac{\dd[2]\xperp \dd[2]\yperp \dd[2]\bperp}{(2\pi)^2} e^{-i\kperp \cdot (\xperp-\yperp)} S(\xperp -\yperp) \\
\times  \left[\frac{1}{(\xperp-\bperp)^2} \ln\frac{\kperp^2 (\xperp-\bperp)^2}{c_0^2} -\frac{(\xperp-\bperp) \cdot (\yperp-\bperp)}{(\xperp-\bperp)^2(\yperp-\bperp)^2} \ln\frac{\kperp^2 \abs{\xperp-\bperp}\abs{\yperp-\bperp}}{c_0^2}\right]\end{multlined}\notag \\
&=\lim_{\rho\to 0}\frac{\alphas \Nc \targetarea}{2\pi}\int \frac{\dd[2]\rperp}{(2\pi)^2} e^{-i\kperp \cdot \rperp} S(\rperp )\left[-\biggl(\ln\frac{\kperp^2}{\rho^2}\biggr)^2+4\int_\rho^\infty\frac{\dd\lperp}{\lperp}\ln \frac{\kperp^2}{\lperp^2}\besselJzero(\lperp \rperp)\right]\notag \\
&= -\frac{\alphas \Nc \targetarea}{4\pi^2}\int \frac{\dd[2] \rperp  }{2\pi} e^{-i\kperp \cdot \rperp} S(\rperp) \biggl(\ln\frac{\kperp^2 \rperp^2}{c_0^2}\biggr)^2, \label{l1}
\end{align}
where we have used $\rho$ as an infrared cutoff to compute the above integration. It is important to notice that the above result is independent of regularization scheme, since the whole expression is finite. Furthermore, by using the following trick (see also Ref.~\cite{Kovchegov:2012mbw}), 
we can evaluate $L_1(\kperp)$ analytically. Let us define 
\begin{equation}
I(\beta)\defn \int_0^\infty \dd\rperp \rperp^{1+\beta} \besselJzero(\kperp \rperp) \exp \left(-\frac{\satscale^2 \rperp^2}{4} \right) =\frac{2^{1+\beta}\Gamma [1+\frac{\beta}{2}] L\bigl[-1-\frac{\beta}{2}, -\frac{\kperp^2}{\satscale^2}\bigr]}{\satscale^{2+\beta}}\, ,
\end{equation}
where $L[-1-\frac{\beta}{2}, x]$ is defined as the Multivariate Laguerre
Polynomial. It is then trivial to find that 
\begin{equation}
L_{q1}(\kperp)=-4\left[I''(0)+2\ln \frac{\kperp}{c_0} I'(0)+\biggl(\ln \frac{\kperp}{c_0}\biggr)^2 I(0)\right] \, ,
\end{equation}
which gives $L_{q1}(\kperp)$ found above. It is useful to note that $L[-1, x]=e^x$ and its first derivative on its first argument $L^{(1,0)}[-1, x]=-\left[\gamma_E+\Gamma[0,x]+\ln x\right]e^{x}$.

Now let us further evaluate $L_1(\kperp)$ in the limit of $\kperp\to \infty$. We can rewrite the integral in question as follows
\begin{multline}
-\frac{1}{\kperp^2}\int_0^\infty \dd z\, z \besselJzero(z) \exp \left(-\frac{\satscale^2}{4\kperp^2} z^2\right) \biggl(\ln\frac{z^2}{c_0^2}\biggr)^2 \\
=\frac{-4}{\kperp^2}\lim_{\lambda\to 0}\int_0^\infty \dd z\, z \besselJzero(z) \exp \left(-\frac{\satscale^2}{4\kperp^2} z^2\right) \left[K_0(\lambda)-K_0(\lambda \frac{z}{c_0})\right]^2.
\end{multline}
It is straightforward to show that the leading power expansion of the above integral gives $\frac{8\satscale^2}{\kperp^4}$.

\section{Some important Fourier transforms}
Although the original derivation~\cite{Chirilli:2011km, Chirilli:2012jd}  of the NLO correction for hadron productions in $pA$ collisions are completed in the coordinate space, in order to achieve better numerical accuracy, we have used Fourier transforms to convert most of the NLO corrections into momentum space in Ref.~\cite{Stasto:2013cha} when the first version of the SOLO package was developed (see \cite{Zaslavsky:2014asa} for more details of the implementation). We only evaluate a couple of NLO terms, for example $\mathcal{H}^{(1)}_{2qq}$, in coordinate space, since the evaluation can be done pretty accurately even in the coordinate space for forward rapidity kinematical region at RHIC and the LHC. 

However, when we try to compare to LHC data at middle rapidity, those terms which are left in the coordinate space, suddenly give huge numerical uncertainty. The typical integration which poses a challenge to numerical integrations is 
\begin{equation}
\int \frac{\dd[2] \xperp}{(2\pi)^2}S(\xperp)\ln\frac{c_0^2}{\xperp^2 \mu^2}e^{-i\kperp\cdot \xperp}.
\end{equation} 
In the LHC middle rapidity kinematical region, the allowed region of $\kperp$ is quite large. Note that $\kperp\equiv \frac{\pperp}{z}$, which is the transverse momentum of the produced parton, can be much larger than $\pperp$ which is the transverse momentum of the measured hadron. Although the expression is well-defined analytically, the above integration oscillates too fast when $\kperp$ is large, therefore it causes a lot of numerical uncertainty. Luckily, we manage to find a way to convert the above integration into momentum space which is much more stable. Using the identity 
\begin{equation}
\int\frac{\dd[2] \xperp}{(2\pi)^2}\ln\frac{c_0^2}{\xperp^2 \mu^2}e^{-i\kperp\cdot \xperp}=\frac{1}{\pi}\left[\frac{1}{\kperp^2}-2\pi\delta^{(2)}(\kperp)\int_0^{\infty}\frac{\dd\lperp}{\lperp}\besselJzero(\frac{c_0}{\mu}\lperp)\right], 
\end{equation}
we find 
\begin{equation}
\int \frac{\dd[2] \xperp}{(2\pi)^2}S(\xperp)\ln\frac{c_0^2}{x_\perp^2 \mu^2}e^{-ik_\perp\cdot x_\perp}=\frac{1}{\pi}\int\frac{d^2 l_\perp}{l_\perp^2}\left[F(k_\perp +l_\perp)-J_0(\frac{c_0}{\mu}l_\perp)F(k_\perp)\right].
\end{equation} 
It is straightforward to test the above identity and find the momentum space expression stable and finite. 

In addition, for the same reason, the new $L_{q1}$ contribution that we found in this paper, as shown in Eq.~(\ref{q2log}), is even more unstable in the coordinate space at the LHC kinematical region. Again, we manage to find a nice trick to convert that expression into momentum space. 
Using dimensional regularization in the $\overline{\textrm{MS}}$ scheme, we can find the following results
\begin{align}
 \int\frac{\dd[2]\lperp}{(2\pi)^2 \lperp^2} e^{-i\lperp\cdot\Rperp} \ln\frac{\kperp^2}{\lperp^2}
 &= \frac{1}{4\pi}\biggl[\frac{1}{\epsilon^2} - \frac{1}{\epsilon}\ln\frac{\kperp^2}{\mu^2} + \frac{1}{2}\biggl(\ln\frac{\kperp^2}{\mu^2}\biggr)^2 - \frac{1}{2}\biggl(\ln\frac{\kperp^2 \Rperp^2}{c_0^2}\biggr)^2 - \frac{\pi^2}{12}\biggr]\label{eq:dimreg1logfourier} \\
 \eval{\int\frac{\dd[2]\lperp}{(2\pi)^2\lperp^2}\ln\frac{\kperp^2}{\lperp^2}}_{\lperp^2\leq\kperp^2}
 &= \frac{1}{4\pi}\biggl[\frac{1}{\epsilon^2} - \frac{1}{\epsilon}\ln\frac{\kperp^2}{\mu^2} + \frac{1}{2}\biggl(\ln\frac{\kperp^2}{\mu^2}\biggr)^2 - \frac{\pi^2}{12}\biggr] , \label{eq:dimreg1log}
\end{align}
The derivation of the first expression can be found in Eq.~(A4) of Ref.~\cite{Mueller:2013wwa}, while the second expression can be computed directly. This trick is inspired by the computations of the Sudakov factors in the saturation formalism. 
Subtracting Eq.~\eqref{eq:dimreg1log} from Eq.~\eqref{eq:dimreg1logfourier} and taking $\epsilon\to 0$ gives the identity
\begin{equation}
 \biggl(\ln\frac{\kperp^2\Rperp^2}{c_0^2}\biggr)^2 = 8\pi\int\frac{\dd[2]\lperp}{(2\pi)^2\lperp^2}\ln\frac{\kperp^2}{\lperp^2}\Bigl[\theta(\kperp - \lperp) - e^{-i\lperp\cdot\Rperp}\Bigr].
\end{equation}
At the end of the day, one can easily find 
\begin{equation}
\int\frac{\dd[2] \rperp}{(2\pi)^2}S(\rperp)\left(\ln\frac{\rperp^2 \kperp^2}{c_0^2}\right)^2e^{-i\kperp\cdot \rperp}=\frac{2}{\pi} \int\frac{\dd[2] \lperp}{\lperp^2} \ln\frac{\kperp^2}{\lperp^2}\left[\theta(\kperp-\lperp)F(\kperp)-F(\kperp+\lperp)\right].
\end{equation}
Up to this point, we have transformed all the NLO corrections into momentum space in SOLO package, which is relatively more stable numerically at both RHIC and the LHC kinematical region.

\end{document}